\documentclass{article}
\usepackage{frascatiphys,here,graphicx,subfigure}
\usepackage{amsmath,amssymb}
\begin{document}
\title{ 
Flavour in Soft Leptogenesis\thanks{\;\;Work done in collaboration with M.~C.~Gonzalez-Garcia,
Enrico Nardi and Juan Racker. Based on hep-ph/1004.5125.}
}
\author{
Chee Sheng Fong       \\
{\em C.~N.~ Yang Institute for Theoretical Physics, Stony Brook University} \\
%
%
%
}
\maketitle
\baselineskip=11.6pt
\begin{abstract}
Successful Soft Leptogenesis (SL) requires a relatively low mass scale for
the SU(2) singlet neutrinos of $10^5-10^8$ GeV. 
However, conventional SL (unflavoured) requires an unnaturally 
small soft supersymmetry(SUSY)-breaking bilinear 
$B \ll \mathcal{O}({\rm TeV})$ coupling for successful leptogenesis. 
On the other hand, in this regime, the interactions mediated by $\tau$, $\mu$ (and even $e$) 
charged lepton Yukawa interactions are in equilibrium, making 
the lepton number asymmetries and the washouts flavour dependent. 
Hence, it is crucial to take into account the flavour effects.
Considering a general soft SUSY-breaking trilinear $A$ couplings, 
it is possible to enhance the efficiency up to $\mathcal{O}(1000)$ compared to the unflavoured case. 
With the enhanced efficiency, we can raise the $B$ up to TeV scale
for successful leptogenesis. Taking into account the low energy constraints, 
we verify that the fast lepton flavour violation processes induced by the
soft SUSY-breaking slepton masses would not destroy the enhancement.
\end{abstract}
\baselineskip=14pt
\section{Soft Leptogenesis Lagrangian and CP Asymmetries}
\label{sec:lag}

The type-I SUSY seesaw model can be described by the superpotential:
\begin{equation}
W=W_{\rm MSSM}+\frac{1}{2}M_{ij}\hat{N}_{i}\hat{N}_{j}+Y_{ik}
\epsilon_{\alpha\beta}\hat{N}_{i}\hat{L}_{k}^{\alpha}\hat{H}^{\beta},
\label{eq:superpotential}
\end{equation}
where $W_{\rm MSSM}$ is the superpotential for the Minimal Supersymmetric Standard Model (MSSM), 
$k=1,2,3$ are the lepton flavour indices,
$M_{ij}$ are the Majorana masses of the right-handed singlet neutrinos with generation indices $i,j$,
and $\hat{N}_{i}$, $\hat{L}_{k}$, $\hat{H}$ 
are the chiral superfields for the right-handed singlet neutrinos, 
the left-handed lepton doublets and the Higgs doublets, 
with $\epsilon_{\alpha\beta}=-\epsilon_{\beta\alpha}$ and $\epsilon_{12}=+1$.

The relevant soft SUSY-breaking terms are given by
\begin{eqnarray} 
-\mathcal{L}_{soft} 
& = & A Z_{ik}\epsilon_{\alpha\beta}\widetilde{N}_{i}\tilde{\ell}_{k}^{\alpha}h^{\beta}
+\frac{1}{2}B_{ij}M_{ij}\widetilde{N}_{i}\widetilde{N}_{j}
+m_{\widetilde{\ell}_{kl}}\widetilde{\ell}_k\widetilde{\ell}_l
+\mbox{h.c.} 
\label{eq:soft_terms}
\end{eqnarray}

The singlet sneutrino and anti-sneutrino states mix, giving rise to the mass
eigenstates:
\begin{eqnarray}
\widetilde{N}_{+i} & = &
\frac{1}{\sqrt{2}}(e^{i\Phi/2}\widetilde{N}_{i}+e^{-i\Phi/2}
\widetilde{N}_{i}^{*}),\nonumber
\\ \widetilde{N}_{-i} & = &
\frac{-i}{\sqrt{2}}(e^{i\Phi/2}
\widetilde{N}_{i}-e^{-i\Phi/2}\widetilde{N}_{i}^{*}),
\label{eq:mass_eigenstates}
\end{eqnarray}
where $\Phi\equiv\arg(BM)$, that correspond to the mass eigenvalues
\begin{eqnarray}
M_{ii\pm}^{2} & = & M_{ii}^{2} \pm|B_{ii}M_{ii}|.
\label{eq:mass_eigenvalues}
\end{eqnarray}

For simplicity, we will concentrate on SL 
arising from a single sneutrino generation $i=1$ and in
what follows we will drop that index.
After superfield phase rotations, we have
three independent physical phases, they are 
\begin{eqnarray}
\phi_{Ak}&=&{\rm arg}(Z_{k} Y^*_k  A B^*), \qquad (k=1,2,3)
\label{eq:CPphase1}
\end{eqnarray}

Eq. (\ref{eq:soft_terms}) leads to CP asymmetries $\epsilon_{k}(T)$ 
arising from self-energy diagrams induced by the bilinear $B$ term,
\begin{eqnarray}
\epsilon_{k}\left(T\right) 
& = & - P_{k}  \, \frac{Z_k}{ Y_k }
\sin\phi_{Ak}\frac{A}{M}\frac{4B\Gamma}{4B^{2}+\Gamma^{2}}
\Delta_{BF}\left(T\right),\;\;\;\;
P_k \equiv  \frac{ Y_{k}^2}{\displaystyle \sum_j Y_{j}^2},
\label{eq:CP_asymres} 
\end{eqnarray}
where
\begin{eqnarray}
\Delta_{BF}(T) & = & 
\frac{c^{s}(T) - c^{f}(T)}{c^{s}(T) + c^{f}(T)}, 
\end{eqnarray}
is the thermal factor associated to the difference between the
phase-space, Bose-enhancement and Fermi-blocking factors for the scalar and fermionic channels, 
that vanishes in the zero temperature limit $\Delta_{BF}(T\!=\!0)=0$\cite{soft1,soft2}. 

\section{Flavour Structure}
Regarding the flavour structure of the soft terms relevant 
for flavoured SL, we can distinguish two general possibilities:

1. {\it Universal soft SUSY-breaking terms}. This case is
realized in supergravity and gauge mediated SUSY-breaking models
(neglecting the renormalization group running of the parameters), 
in our notations corresponds to
\begin{equation}
Z_k=  Y_k. 
\label{eq:uts}
\end{equation}
In this case the only flavour structure arises from the Yukawa couplings and both the CP
asymmetries $\epsilon^k$ and the corresponding washout terms are proportional 
to the same $P_k$, resulting in mild enhancement of ${\cal O }(30)$
in efficiency from one-flavour approximation \cite{oursoft}. 
We refer to this case as {\sl Universal Trilinear Scenario} (UTS).

2. {\it General soft SUSY-breaking terms}. The most general form for 
the soft-SUSY breaking terms is allowed, only subject to the 
phenomenological constraints from limits on flavour 
changing neutral currents (FCNC) and from lepton flavour violating 
(LFV) processes. To simplify the analysis while still capture
some of the main features of the general case, we choose
\begin{equation}
Z_{k}=\frac{\displaystyle
\sum_j |Y_{j}|^2}{3 Y^*_{k}}\; , 
\label{eq:ems}
\end{equation}
such that the CP asymmetries become flavour independent,
$\epsilon^k =\epsilon/3$ for each flavour.
In what follows we will refer to this case as 
the {\sl Simplified Misaligned Scenario} (SMS). 
With this choice, since it is possible to reduce the washout in
a particular flavour direction while keeping the corresponding CP asymmetry fixed,
a much greater enhancement than the UTS becomes possible.

In both the UTS and SMS, from (\ref{eq:CPphase1}), we see that we only have
one unique phase $\phi_A={\rm arg}(A B^*)$.

\section{Results}
We numerically solved the relevant Boltzmann Equations
\footnote{For more details, please refer to Ref.~\cite{oursoft2}}
to obtain the final amount of ${B}-{L}$ asymmetry generated in the decay of the 
singlet sneutrinos (assuming no pre-existing
asymmetry) which can be parametrized as:
\begin{equation}
Y_{B-L}(z\rightarrow \infty)=
-2 \eta\, \bar\epsilon \, Y^{\rm eq}_{\widetilde N}(T\gg  M),\;\;\;\;\eta\equiv\sum_k \eta_k.
\label{eq:yb-l}
\end{equation}

After conversion by the sphaleron transitions, the final baryon asymmetry
is related to the ${B}-{L}$ asymmetry by
\begin{equation}
Y_{B}=\frac{8}{23} \, Y_{B-L}(z\rightarrow \infty). 
\label{eq:yb}
\end{equation}

We also define the $\tilde N_\pm$ decay parameter, $m_{\rm eff}
\equiv \sum_k Y^2_k v_u^2/M$ which is related to the washout parameter $K$ as
$K=\Gamma_{\widetilde{N}}/H(M)=m_{\rm eff}/m_*$ where 
$\Gamma_{\widetilde{N}}$ is the total singlet sneutrino decay width,
$v_u = v \sin\beta$ (with $v=174$ GeV),  
$m_*=\sqrt{\frac{\pi g^*}{45}}\times\frac{8\pi^2v^2_u}{m_P}\sim 10^{-3}\,\,{\rm eV}$ 
with $g_{s}^{*}$ the total number of relativistic degrees of freedom ($g^{*}=228.75$ in the MSSM)
and $m_P$ the Planck mass.

\begin{figure}[H]
    \begin{center}
        {\includegraphics[scale=0.34]{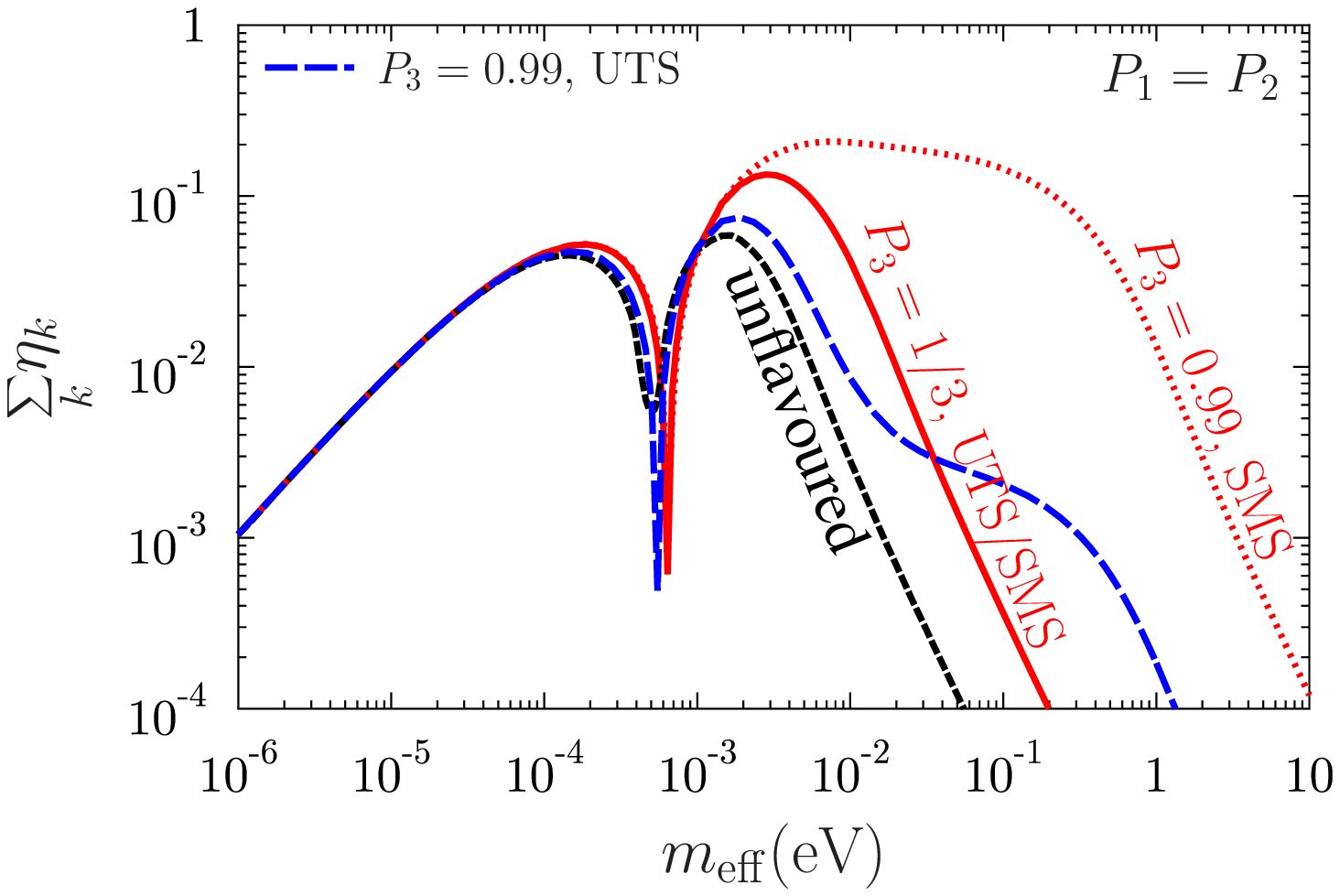}}
        {\includegraphics[scale=0.34]{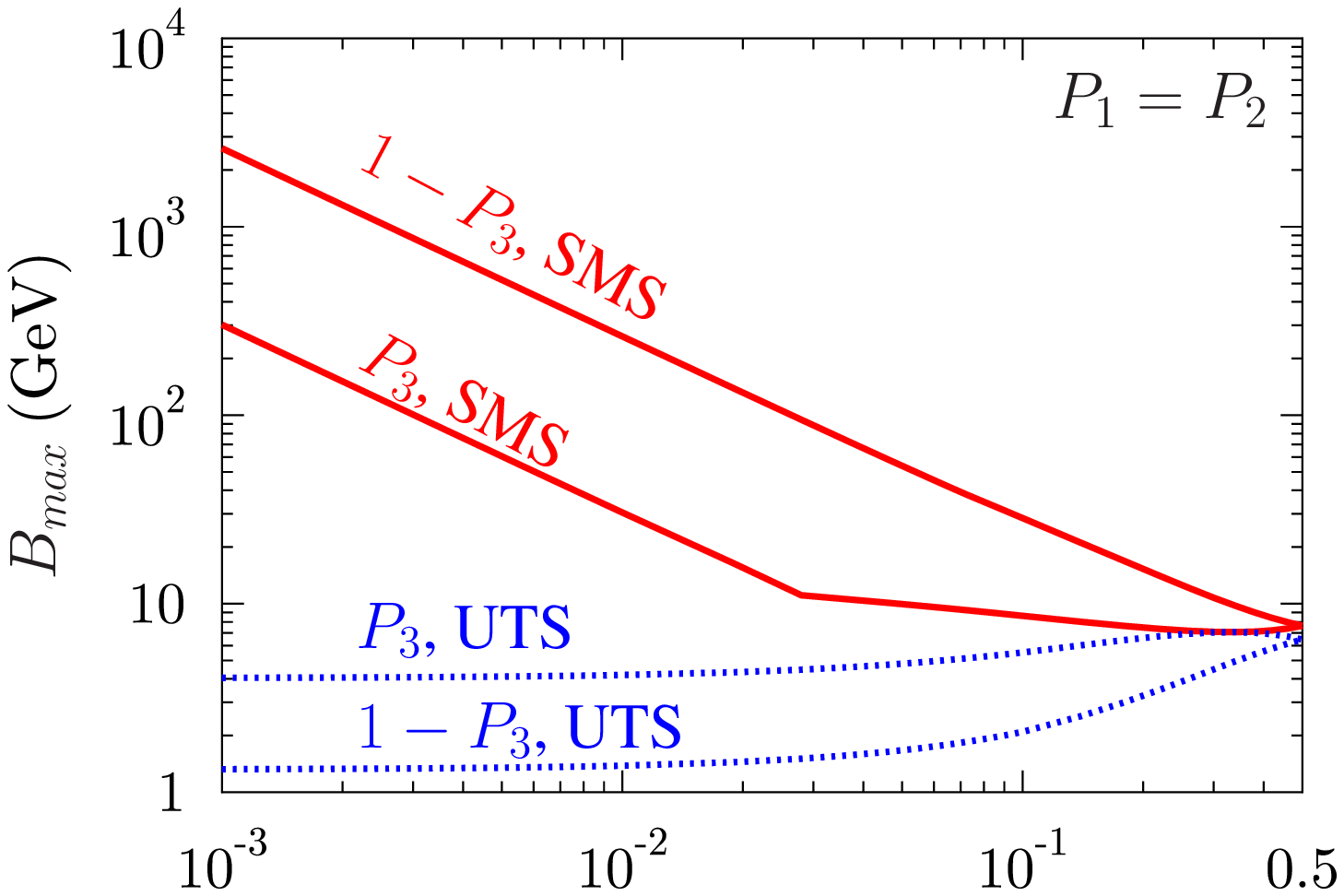}}
        \caption{
Left: The dependence of the efficiencies on $m_{\rm eff}$.
Right: Maximum values of $B$ which can lead to successful leptogenesis 
as a function of $P_3$ and $1-P_3$.  
The figures correspond to $A\sin\phi_A$=1 TeV and $\tan\beta=30$.}
\label{fig:efficiency}
    \end{center}
\end{figure}

In the left panel of Fig.~(\ref{fig:efficiency}), we plot the efficiencies as a
function of $m_{\rm eff}$ for both the scenarios UTS and SMS.
Deviating from the flavour equipartition case $P_1=P_2=P_3=1/3$,
in the SMS, we can obtain an enhancement up to $\mathcal{O}(1000)$ 
compared to the one-flavoured approximation. With this enhancement,
it is possible to push the values of $B$ up to natural values at TeV scale for
successful leptogenesis as shown in the right panel of Fig.~(\ref{fig:efficiency}). 

\vspace{-8mm}

\section{Low Energy Constraints}
\begin{figure}[H]
    \begin{center}
        {\includegraphics[scale=0.37]{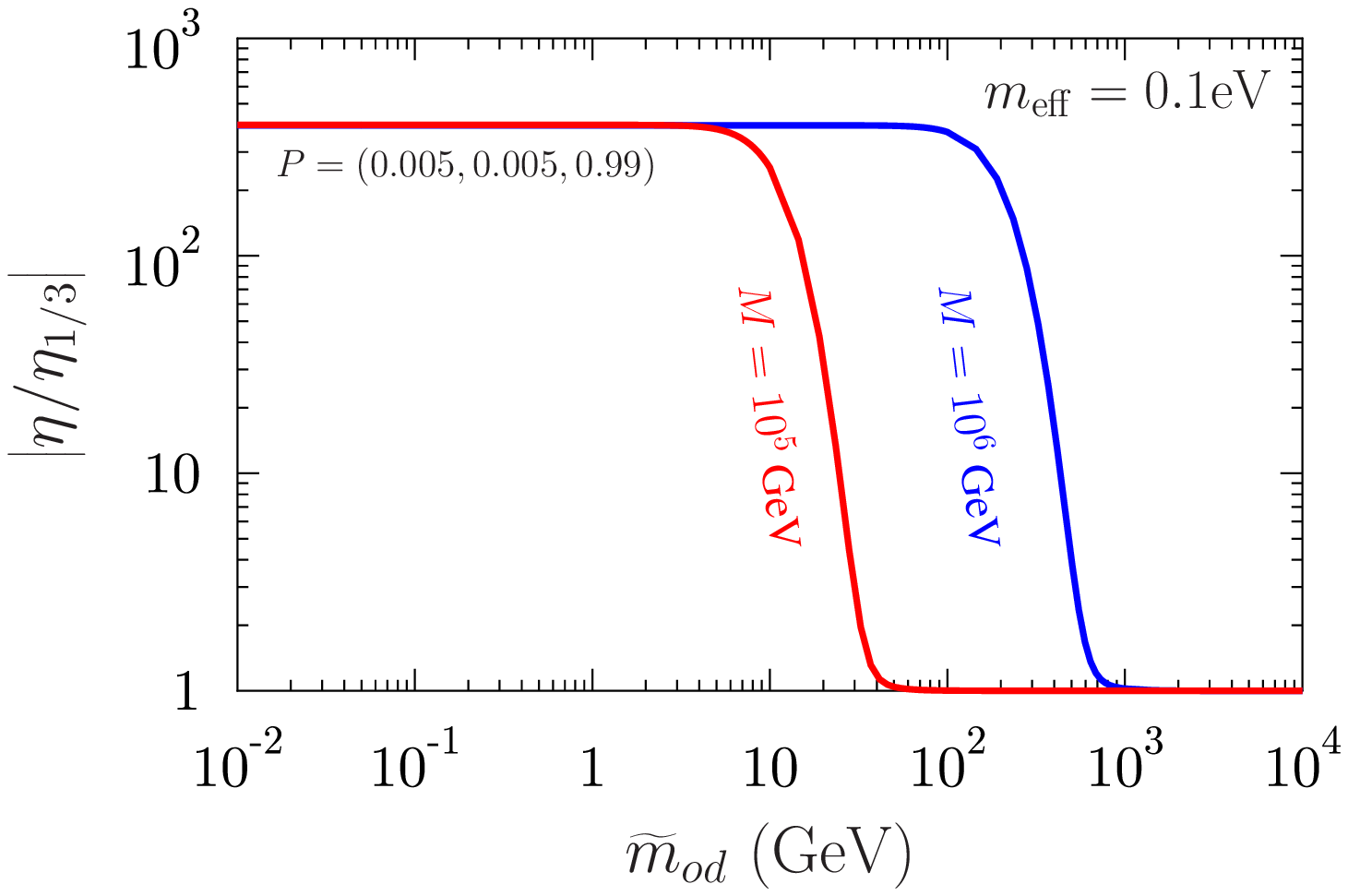}}
        {\includegraphics[scale=0.34]{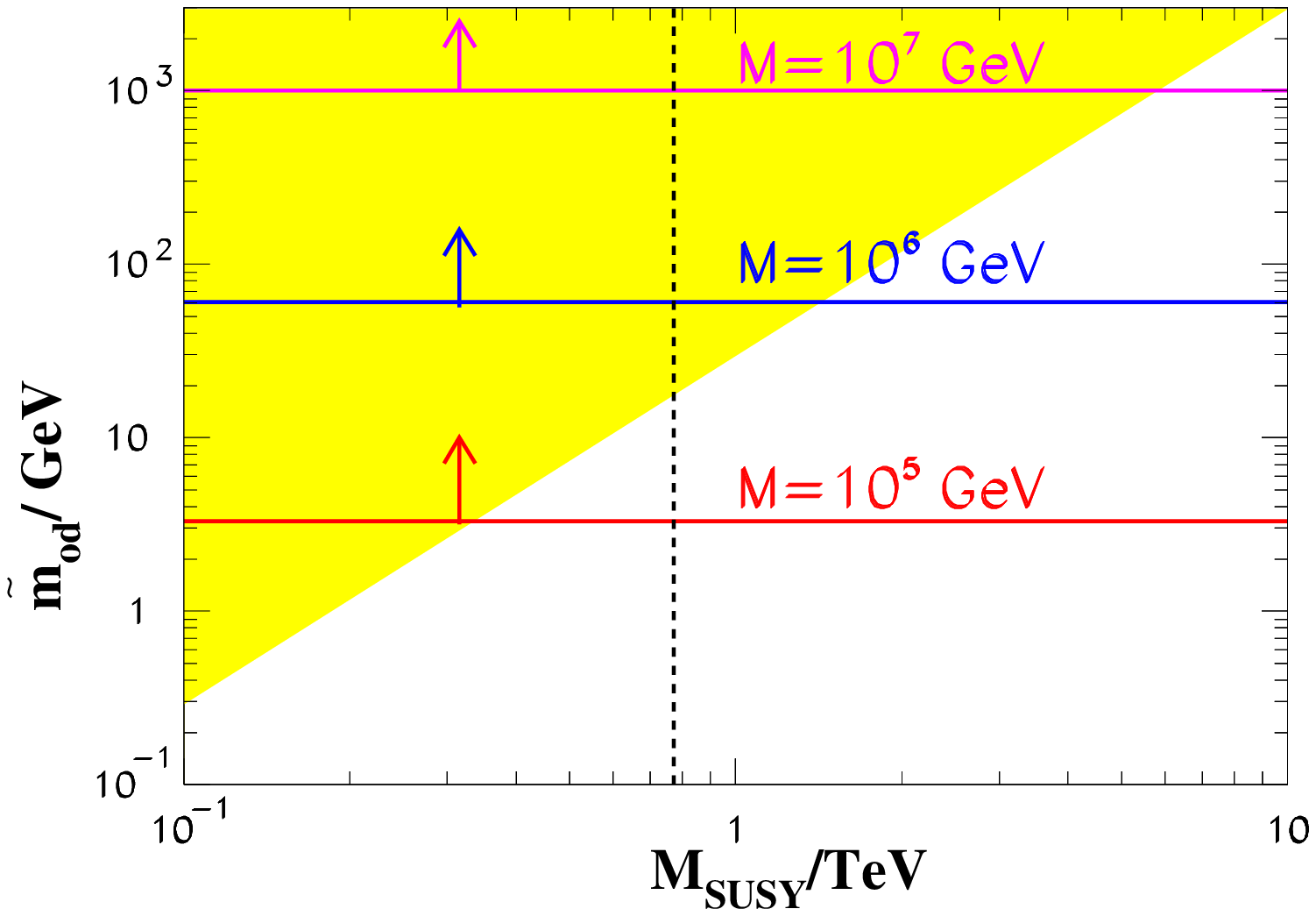}}
      \caption{
  Left: The dependence of the efficiencies (normalized to the flavour
  equipartition case) on the off-diagonal soft slepton mass parameter $\widetilde{m}_{od}$.
  Right: Excluded region (shaded in yellow) of
  $\widetilde{m}_{od}$ versus $m_{\rm SUSY}$ arising
  from the present bound of $BR(\mu\rightarrow e\gamma)\leq 1.2 \times
  10^{-11}$, together with the minimum value of
  $\widetilde{m}_{od}$ for which LFE effects start damping
  out flavour effects. Three lines shown corresponds 
  to $M=10^5\,$GeV, $M=10^6\,$GeV and $M=10^7\,$GeV.
  The vertical dashed line represents the value of $m_{\rm SUSY}/(30)^\frac{1}{2}$ 
  required to explain the discrepancy between the Standard Model (SM)
  prediction and the measured value of muon anomalous magectic moment $a_\mu$~\cite{nuria}. 
  In the plots, we assume $\tan\beta=30$.}
\label{fig:LFE}
    \end{center}
\end{figure}

As has been highlighted in Ref.~\cite{lfe}, at sufficiently  
low temperatures the off-diagonal soft-SUSY breaking slepton masses can give rise to  
lepton flavour equilibration (LFE), effectively damping all dynamical 
flavour effects. In the left panel of Fig.~(\ref{fig:LFE}), 
we show the dependence of the efficiencies (normalized to the flavour equipartition case) 
on the off-diagonal soft slepton mass 
parameter ${m}_{od}$ which for simplicity, we assume to be flavour independent.
We see that there is a cut-off value of $\widetilde{m}_{od}$ for each $M$ such that the enhancement is totally 
damped out. In the right panel of Fig.~(\ref{fig:LFE}), 
we show that in most part of the SUSY parameter space that is relevant for SL,
subjecting to low energy constraints, 
the large flavour enhancements can survive the LFE effects.

\vspace{-14mm}

\section{Acknowledgements}
The speaker thanks Concha Gonzalez-Garcia, Enrico Nardi and Juan Racker with whom he collaborated with in this work.
He is also grateful to the 2\textsuperscript{nd} Young Researchers Workshop 
(in conjuction with XV Frascati Spring School ``Bruno Touschek'') organizing committee for the hospitality during his stay in beautiful Frascati.

\end{document}